\documentclass[aps,prl,floatfix,twocolumn,superscriptaddress,showpacs,amsmath,amssymb]{revtex4}
\usepackage{graphicx}
\usepackage{epstopdf}
\usepackage{epsfig} 
\allowdisplaybreaks
\newcommand{\be}{\begin{equation}}
\newcommand{\ee}{\end{equation}}
\newcommand{\bea}{\begin{eqnarray}}
\newcommand{\eea}{\end{eqnarray}}
\newcommand{\ba}{\begin{eqnarray*}}
\newcommand{\ea}{\end{eqnarray*}}
\newenvironment{eqs}%
{\begin{equation} \begin{aligned}}%
{\end{aligned} \end{equation} }
\newcommand{\beal}{\begin{eqs}}
\newcommand{\eal}{\end{eqs}}
\newcommand{\bR}{\mathbf{R}}

\newcommand{\bRp}{\mathbf{R'}}

\newcommand{\dis}{\displaystyle}
\newcommand{\fract}[2]{\frac{\dis #1}{\dis #2}}
\newcommand{\Tr}{\text{Tr}}
\newcommand{\eqn}[1]{(\ref{#1})}
\newcommand{\ket}[1]{\mid\! #1\rangle}
\newcommand{\bra}[1]{\langle #1\!\mid} 
\newcommand{\bw}{\begin{widetext}}
\newcommand{\ew}{\end{widetext}}
\newcommand{\esp}[1]{\text{e}^{#1}}

\begin{document}

\title{Selective cooling by impulse perturbations in a simple toy model}

\author{Michele Fabrizio} 
\affiliation{International School for
  Advanced Studies (SISSA), Via Bonomea
  265, I-34136 Trieste, Italy} 

\date{\today} 


\begin{abstract}
We show in a simple exactly-solvable toy model that a properly designed 
impulse perturbation can transiently cool down 
low-energy degrees of freedom at the expenses of high-energy ones that heat up. The model consists of two infinite-range quantum Ising models, one, the \textit{high-energy} sector, with a transverse field much bigger than the other, the \textit{low-energy} sector. The finite-duration perturbation is a spin-exchange that couples the two Ising models with an oscillating coupling strength. We find a cooling of the low-energy sector that is optimised by the oscillation frequency in resonance with the spin-exchange excitation. After the perturbation is turned off, the Ising model with low transverse field 
can even develop spontaneous symmetry-breaking despite being initially above the critical temperature.  
\end{abstract}
\maketitle

Crystalline solids, either metallic or insulating, are characterised by electronic and lattice degrees of freedom whose dynamics is controlled by a hierarchy of energy scales and relaxation times 
that are directly accessible by pump-probe spectroscopy\cite{Claudio-Adv,Nicoletti&Cavalleri-AOP2016}. In such experiments, a sample is driven away from equilibrium by an intense laser pulse, the pump, and the subsequent relaxation dynamics is probed by a variety of spectroscopic tools as function of the time delay from the pump pulse, with a resolution that today can well achieve the attosecond\cite{attosecond}. Selected degrees of freedom can be excited by properly tuning the laser frequency, thus offering an unprecedented wide choice of non-equilibrium pathways unaccessible by conventional experiments where thermodynamic state variables, like temperature, pressure or chemical composition, are varied. Moreover, the early-time relaxation dynamics is essentially unaffected by the environment, which starts playing a role only nanoseconds after the pulse. In other words, the environment just determines the initial equilibrium conditions of the sample, which then evolves for a relatively long time as it were effectively isolated. This is just what happens in cold atoms systems\cite{cold-atoms-review}, with the major difference that real materials are more complex, and thus potentially much richer though harder to model. \\
As an example, we here mention a recent pump-probe experiment on K$_3$C$_{60}$ alkali fullerides~\cite{Cavalleri-Nature2016}. When shot by a laser pulse in the mid-infrared, $80 - 200$~\text{meV}, frequency range, this 
molecular conductor, which at equilibrium becomes a superconductor below $T_c\sim 20~\text{K}$, shows a transient, few picoseconds long, superconducting-like optical response that is observed up to temperatures 
$T\sim 200~\text{K}$, ten times higher than $T_c$. Bearing in mind that in K$_3$C$_{60}$, as in many other correlated superconductors\cite{Basov-RMP2011},  the transition to a normal metal occurs by gap filling rather than closing, 
the transient superconducting signal looks as if the laser pulse had swept away the thermal excitations that at equilibrium fill the gap\cite{mio}. In other words, it appears that the light promotes the excitations responsible of the mid-infrared absorption, but, at the same time, it cools down the low frequency, $\lesssim 20~\text{meV}$,  electronic degrees of freedom. An explanation of this \textit{selective cooling} was proposed in Ref.~\cite{mio} based on the prediction that the mid-infrared absorption is due to a localised spin-triplet exciton that can be populated by light through the concurrent absorption/emission of 
spin-triplet particle-hole excitations of the Fermi liquid. Since at equilibrium the thermal population of spin-triplet particle-hole pairs is more abundant than that of the excitons, the net effect is an entropy flow from the former to the latter within the pulse duration. After the pulse ends, that entropy slowly flows back by the non-radiative exciton recombination, which entails the existence of a whole time interval in which the Fermi liquid is effectively cooler than at equilibrium, in agreement with the experimental evidence.\\
\begin{figure}[t]
\centerline{\hspace{-0.5cm}\includegraphics[width=0.55\textwidth]{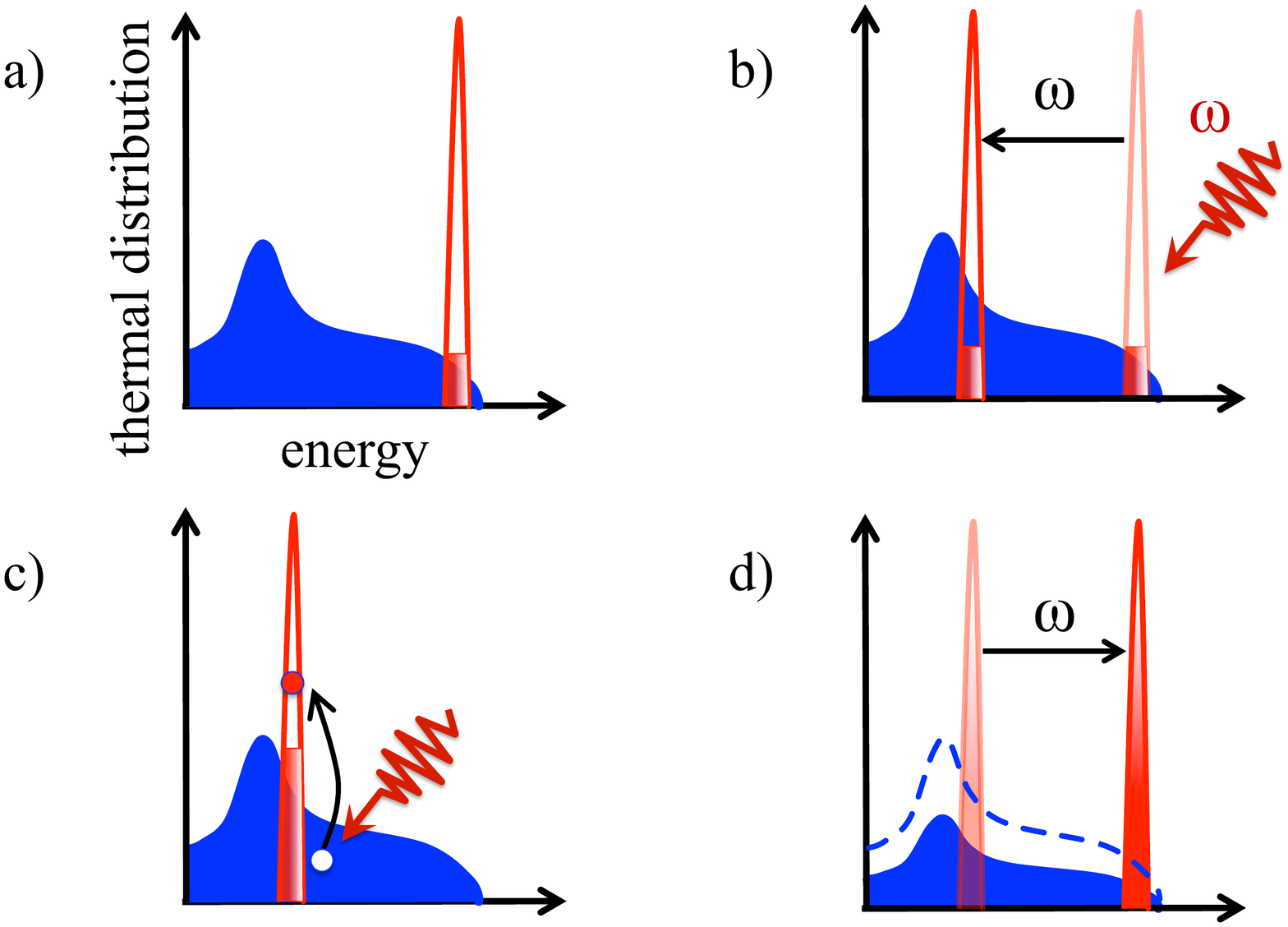}}
\vspace{-0.5cm}
\caption{Pictorial explanation of the cooling mechanism within the rotating-wave approximation (RWA). a) The initial thermal distribution of low-energy (blue) and high-energy (red) excitations; b) a laser pulse with frequency $\omega$ shoots the system. In the RWA the high-energy peak is effectively shifted downward by $\omega$ for the whole pulse duration; c) the amplitude of the laser induces a transfer of thermal excitations to the formerly poorly-populated peak; d) when the laser is turned off,  the peak returns back to its initial position, but it is now overpopulated at the expenses of the low-energy excitations.} 
\label{explain}
\end{figure}
Such cooling mechanism is in fact quite generic\cite{GeorgesPRA2009}, as it requires, in essence, the existence of an entropy sink that opens when the laser is on, while, when the laser is turned off, it gradually gives back the stolen entropy. A simple-minded pictorial explanation within the rotating-wave approximation is drawn in Fig.~\ref{explain}. Here we shall show how to practically achieve such cooling in a simple 
exactly-solvable toy model.\\ 
Specifically, we consider the following unperturbed Hamiltonian 
\be 
H = \sum_{n=1}^2\,H_n\,,\label{Ham}
\ee
where, for each $n=1,2$, 
\beal
H_n &= -\fract{J_n}{2V}\,\sum_{\bR\bRp}\,\sigma^x_{n\bR}\,\sigma^x_{n\bRp} 
- h_n\,\sum_\bR\,\sigma^z_{n\bR}\,,\label{Hn}
\eal
describes an infinite-range quantum Ising model, a special case of the so-called Lipkin-Meshkov-Glick model\cite{Lipkin-1965}, 
$V\!=\!2M$ is the number of sites, while $\sigma^a_{n\bR}$ is the 
$a=x,y,z$ component of the $n=1,2$ Ising spin at site $\bR$. 
For simplicity we shall take $J_1=J_2=1$ as energy unit. \\
The model \eqn{Hn} for each $n=1,2$ admits as conserved quantity the total spin 
with eigenvalue $\text{S}_n(\text{S}_n+1)$, 
where $\text{S}_n = 0,1,\dots,M$. Each energy eigenstate is thus labelled also by a value of $\text{S}_n$ and its degeneracy $g(\text{S}_n)=C^{2M}_{S_n} - C^{2M}_{S_n+1}$, where $C^n_m$ are binomial coefficients, corresponds to the number of ways $2M$ spin-1/2's can give a total spin $\text{S}_n$. If we set 
$\text{S}_n = M N_n$, then, in the thermodynamic limit $M\to\infty$, $N_n$  effectively becomes a continuous variable $\in[0,1]$ and the partition function can be evaluated semiclassically\cite{Lieb,Sengupta-PRB2006,Bapst&Semerjian-2012}
\bea
&&\!\!\!Z_n = \text{e}^{-\beta F_n}\!=\!\sum_{\text{S}_n}\,g(\text{S}_n)\!\!\!\sum_{\text{S}_n^z=-\text{S}_n}^{\text{S}_n}\!\!
 \bra{\text{S}_n,\text{S}_n^z} \!\esp{-\beta\,H_n}\!\ket{\text{S}_n,\text{S}_n^z} \nonumber\\
&&\;\;\simeq  M\!\int_0^{1}\!\!\!dN_n\big(2\text{S}_n+1\big)\!\int\fract{d\cos\theta_n\,d\phi_n}{4\pi}\,
\esp{-2M\beta f_n},\label{Z_n}
\eea
where the semiclassical free-energy density reads
\be
f_n \!= \!-\fract{N_n^2}{2}\sin^2\theta_n \cos^2\phi_n - h_n N_n \cos\theta_n -T \mathcal{S}\big(N_n\big)\,,
\label{free-energy}
\ee
with the entropy density  
\beal
\mathcal{S}\big(N_n\big) &= \fract{\ln g(\text{S}_n)}{2M}
\simeq \ln 2 -\big(1+N_n\big)\ln \sqrt{1+N_n}\\
& \qquad 
-\big(1-N_n\big)\ln \sqrt{1-N_n}\,,\label{entropy}
\eal
which vanishes at $N_n =1$, and increases with decreasing $N_n$ up 
to $\ln 2$ at $N_n=0$. In the thermodynamic limit the partition function is dominated by the saddle point of \eqn{free-energy}, which also implies that at any temperature $T$ the Boltzmann density matrix 
becomes a projector onto the ground state $\ket{\text{GS}_n;N_n}$ within the subspace with $N_n=N_n(T)$\cite{Bapst&Semerjian-2012}.
One readily finds\cite{Bapst&Semerjian-2012} that, if 
$h_n\leq J_n=1$ and $T\leq T_{nc}$, where   
\be
T_{nc} = 
 2h_n \Big/ \ln\fract{1+h_n}{1-h_n}
\;,\label{Tnc}
\ee
then $N_n(T)$ is the solution of the equation  
\be
N_n(T) = \tanh\beta\,N_n(T)\geq h_n\,,\label{N(T)-ord}
\ee
and in the ground state $\ket{\text{GS}_n;N_n(T)}$ 
the $Z_2$ symmetry, $\sigma^x_{n\bR} \to - \sigma^x_{n\bR}$, 
$\forall\,\bR$, of the model \eqn{Hn} is spontaneously broken 
with order parameter 
$\langle\,\sigma^x_{n\bR}\,\rangle^2 = N_n(T)^2-h_n^2$,
which corresponds to the Euler angles  
$\theta_n = \cos^{-1}h_n/N_n(T)$ and $\phi_n=0$ in Eq.~\eqn{free-energy}.
Above $T_{nc}$, or if $h_n>1$, the symmetry is restored, i.e. 
$ \langle\,\sigma^x_{n\bR}\,\rangle^2 =0$, $\theta_n=0$, and   
\be
N_n(T) = \tanh \beta\,h_n\,.\label{N(T)-dis}
\ee
The above semiclassical results, which are exact in the thermodynamic limit, can be rederived by a mean-field density matrix $\rho_n=\prod_\bR\,\rho_{n\bR}$, such that $\Tr\big(\rho_{n\bR}\big)=1$,  through the variational principle
\be
F_n \leq \min_{\{\rho_{n\bR}\}}\!\!\bigg(\Tr\Big(\rho_n\,H_n\Big) 
+T\sum_\bR\Tr\Big(\rho_{n\bR}\,\ln\rho_{n\bR}\Big)\bigg)\,.\label{variational-principle}
\ee
The above inequality becomes a true equality in the thermodynamic limit because of the infinite connectivity, 
which implies that, for $\bR\not=\bRp$, $\langle\sigma^a_{n\bR}\,\sigma^b_{n\bRp}\rangle 
- \langle\sigma^a_{n\bR}\rangle\langle\sigma^b_{n\bRp}\rangle \sim 1/M$ and thus vanishes
for $M\to\infty$. Minimising the right hand side of Eq.~\eqn{variational-principle}, one readily 
finds\cite{Supplemental} that    
\beal
\rho_{n\bR}(T) &= Z_{n\bR}^{-1}\;U_{n\bR}\,
\text{e}^{-\beta\,H_{n\bR}(T)}
\;U_{n\bR}^\dagger\,,\label{rho_R}
\eal
where 
$
U_{n\bR} = \exp\Big(\!-i\phi_n\,\sigma^z_{n\bR}/2\Big)
\exp\Big(\!-i\theta_n\,\sigma^y_{n\bR}/2\Big)$,
with the same Euler angles as before, and 
\be
H_{n\bR}(T) = -\mu_{n}(T)\,\sigma^z_{n\bR}\,,\label{HnR}
\ee
so that $Z_{n\bR} = 2\cosh \beta\mu_n(T)$. The effective field 
$\mu_n(T)=N_n(T)$ if 
$h_n\leq 1$ and $T\leq T_{nc}$, otherwise $\mu_n(T)=h_n$. This variational scheme leads to the 
same free energy as in the semiclassical approach, but provides additional useful information, for instance 
that the energy of an excitation that changes $\text{S}_n$ by 
$\pm 1$ is simply $\mp 2\mu_n(T)$. We end mentioning that, even though  the variational density matrix does not commute, as expected, with the total spin operator, nonetheless its relative fluctuation vanish in the thermodynamic limit.\\

\begin{figure}[t]
\vspace{-2cm}
\centerline{\includegraphics[width=0.4\textwidth,angle=-90]{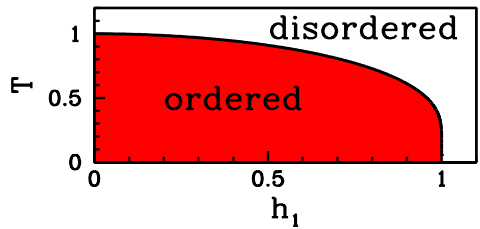}}
\vspace{-1.5cm}
\caption{Phase diagram of the Hamiltonian \eqn{Ham} with 
$h_2=10$ as a function of $h_1$. In the red coloured region below the critical temperature $T_c$ the $Z_2$ symmetry of subsystem 1 is spontaneously broken. } 
\label{phase-diagram}
\end{figure}

\begin{figure}[thb]
\vspace{-0.9cm}
\centerline{\includegraphics[width=0.4\textwidth,angle=-90]{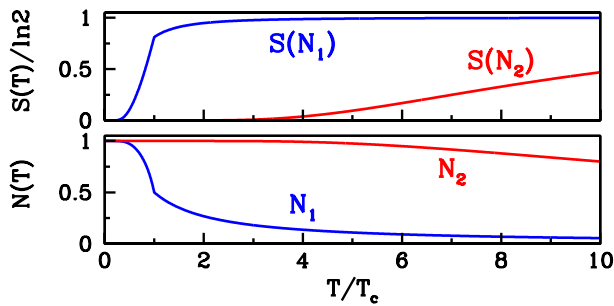}}
\vspace{-1.6cm}
\caption{$N_1$ and $N_2$ (bottom panel) and the corresponding entropies, $\mathcal{S}(N_n)$ (top panel), as function of temperature for $h_1=0.5$ and $h_2=10$.} 
\label{occupations}
\end{figure}

\begin{figure}[bht]
\vspace{-0.5cm}
\centerline{\includegraphics[width=0.4\textwidth,angle=-90]{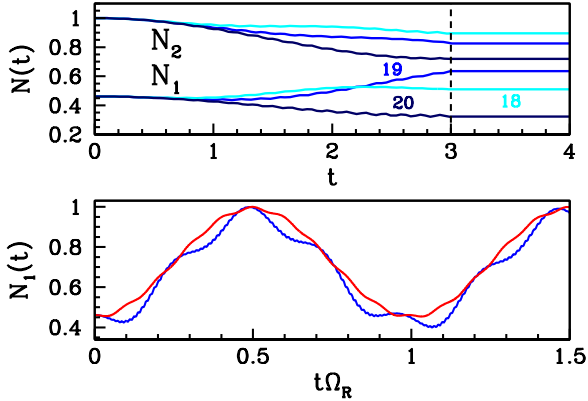}}
\vspace{-.8cm}
\caption{Top panel: time evolution of $N_1(t)$, the three lower curves, and $N_2(t)$, the three upper ones, 
for $h_1=0.5$, $h_2=10$, $T=1.1\,T_c$, $\tau=3$, $E_0=0.4$ and $\omega=18$, cyan lines, 
$\omega=19$, blue, and finally $\omega=20$, dark blue. We emphasise that beyond $\tau$, the vertical dashed line, $N_n(t)$ remain constant.
Bottom panel: time evolution of $N_1(t)$ with the same $h_1$, $h_2$ and $T$ as before, $\omega=\Delta$, but now with a large $\tau$ and two different values of $E_0$: $E_0=0.5$, blue, and $E_0=0.2$, red. We note clear Rabi oscillations, with $\Omega_\text{R}= E_0/2\pi$ the Rabi frequency. } 
\label{Nt}
\end{figure}

\begin{figure}[t]
\vspace{-0.7cm}
\centerline{\includegraphics[width=0.4\textwidth,angle=-90]{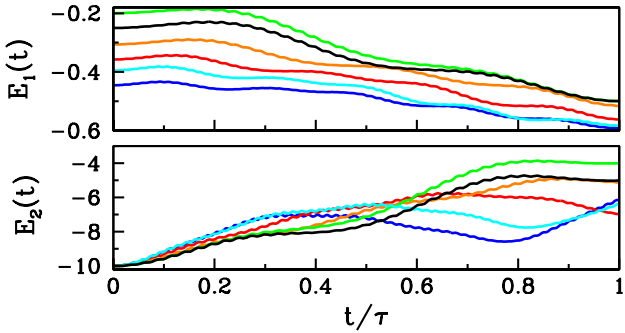}}
\vspace{-1.6cm}
\caption{Time evolution of the internal energy $E_1(t)$ of subsystem 1, top panel, and $E_2(t)$ of subsystem 2, bottom panel, at $h_1=0.5$, $h_2=10$, 
$E_0=0.5$, the frequency $\omega=\Delta$ at resonance, and different temperatures, $T/T_c=1.2,1.0,0.94,
0.90,0.86,0.80$, from the upper to the lower curve in the top panel. In the bottom panel the colours correspond to the same $T$ as in the top one.} 
\label{Et}
\end{figure}

\noindent
Coming back to the full unperturbed Hamiltonian \eqn{Ham}, 
we can for a start conclude that the Boltzmann density matrix can be well approximated by 
\be
\rho=\prod_\bR\,\rho_{1\bR}(T)\,\rho_{2\bR}(T)\,, \label{rho_12}
\ee
with $\rho_{n\bR}(T)$ of Eq.~\eqn{rho_R}.   
Since our aim is to describe two coupled systems with degrees of freedom well separated in energy, we choose as Hamiltonian parameters $h_1<1$ and $h_2\gg 1$, so that $\mu_2(T)\gg\mu_1(T)$, see
Eq.~\eqn{HnR}, and the subsystem 2 is disordered at any temperature. The phase diagram is shown 
in Fig.~\ref{phase-diagram}.  
In Fig.~\ref{occupations} we plot the temperature dependence of 
$N_1$ and $N_2$, see equations \eqn{N(T)-ord} and \eqn{N(T)-dis}, and of the corresponding entropies $\mathcal{S}(N_n)$, 
Eq.~\eqn{entropy}, for $h_1=0.5$ and $h_2=10$. We note the existence of a wide temperature range, which extends from well below $T_c$ to well above it, where subsystem 2 is poor in 
entropy, $\mathcal{S}(N_2)\ll 1$, unlike subsystem 1, $\mathcal{S}(N_1)\lesssim \ln 2$. 
For our purposes this is a favourable circumstance, as we can exploit subsystem 2 as entropy sink of subsystem 1. We therefore assume that initially, $t= 0$, the system is prepared into the thermal state at temperature $T$ and 
then, from $t=0$ until $t=\tau$ the Hamiltonian changes into 
$H+\delta H(t)$, where the perturbation 
\be
\delta H(t) = - E_0\,\cos\omega t\,\sum_{\bR}\,\sigma^x_{1\bR}\,
\sigma^x_{2\bR}\,,\label{deltaH}
\ee
mimics a laser pulse of duration $\tau$, frequency $\omega$ and peak amplitude $E_0$. Above 
$t=\tau$ the perturbation is turned off and the system evolves unitarily, again with Hamiltonian $H$ in \eqn{Ham}. 
In presence of the perturbation, $\text{S}_1=M N_1$ and $\text{S}_2=M N_2$ are not anymore conserved quantities, since $\delta H(t)$ changes both 
$\text{S}_1$ and $\text{S}_2$ by $\pm 1$. If $T$ is such that $N_2\simeq 1\gg N_1$, see Fig.~\ref{occupations}, we expect that in the early stage 
$\delta H(t)$ lowers $\text{S}_2$ while raises $\text{S}_1$ through a sequence of 
elementary excitations $\text{S}_2\to \text{S}_2-1$ and $\text{S}_1 \to \text{S}_1+1$, each with energy cost $\Delta = 2\mu_2(T) - 2\mu_1(T)$, see Eq.~\eqn{HnR}.   
Therefore, after the pulse stops and if $\tau$ is not too large, a net flow of entropy has occurred from subsystem 1 to 2, which thus corresponds to an effective cooling of the former and corresponding heating of the latter. \\
To confirm this expectation, we study the time evolution $\rho(t)$ of the density matrix \eqn{rho_12}, which, because of the infinite connectivity, can be still written as $\rho(t) = \prod_\bR\,\rho_\bR(t)$, where 
$\rho_\bR(t)$ satisfies the equation of motion ($\hbar=1$) 
$
i\,\dot{\rho}_{\bR}(t) = \Big[H_\bR(t)\,,\,\rho_\bR(t)\Big]
$, 
with boundary condition $\rho_\bR(t=0) = \rho_{1\bR}(T)\,\rho_{2\bR}(T)$, and where  
\[
H_\bR(t)\! =\! -\sum_{n=1}^2\Big(h_n\,\sigma^z_{n\bR} + m_n(t)\,\sigma^x_{n\bR}\Big) 
\! - \!E(t)\,\sigma^x_{1\bR}\,
\sigma^x_{2\bR}\,,\label{H_R(t)}
\]
being $E(t)=E_0\cos\omega t$ if $0< t\leq \tau$, and $E(t)=0$ otherwise. The time-dependent fields 
$m_n(t)$, $n=1,2$,  are determined by the  self-consistency condition 
$
m_n(t) = \Tr\Big(\rho_\bR(t)\,\sigma^x_{n\bR}\Big)$, $\forall\,\bR
$.
The equations of motion can be numerically integrated with no particular difficulty\cite{Supplemental}, and therefore we here quote just the outcomes. \\
In the top panel of Fig.~\ref{Nt} we show $N_1(t)$ and $N_2(t)$ for 
$h_1=0.5$, $h_2=10$, $T = 1.1 \,T_c$, $\tau=3$, $E_0=0.4$, and three different frequencies $\omega=18,19,20$. We observe that the maximum increase of $N_1(\tau)$ occurs when $\omega=19$, while for $\omega=18$ the increase is lower and for $\omega=20$ we even find a decrease. This is not surprising since 
$\omega=19$ is exactly in resonance with the excitation that lowers $\text{S}_2$ by one and concurrently increases $\text{S}_1$ by the same amount, and which indeed costs 
energy $\Delta=2\mu_2(T)-2\mu_1(T)=2h_2-2h_1=19$.  
Hereafter we shall thus stick to resonance, $\omega = \Delta$. In the bottom panel of Fig.~\ref{Nt} we instead plot $N_1(t)$ for a large value of $\tau$, the same $h_1$ and $h_2$ as before, and two different values of $E_0=0.5,0.2$. 
The time is shown in units of the inverse Rabi frequency $\Omega_\text{R}= E_0/2\pi$. We clearly observe Rabi oscillations, so that, if our aim is to fix the pulse duration so as to get the maximum increase of 
$N_1(t)$ in the minimum $\tau$, then the best choice is $\tau$ around half of the Rabi period, which we shall adopt in what follows. \\
In Fig.~\ref{Et}, top panel, we show the energy density
$E_1(t) = \Tr\big(\rho(t)\,H_1\big)/V$ for different initial temperatures, from above to below $T_c$.  Recalling that $E_1(t)$ remains constant for $t\geq \tau$, the results shown \textit{imply a significant cooling down of subsystem 1 after the pulse ends}. This is evidently counterbalanced by a concurrent heating of subsystem 2, 
as shown in the bottom panel of Fig.~\ref{Et}. 

\begin{figure}[t]
\vspace{-0.4cm}
\centerline{\includegraphics[width=0.5\textwidth,angle=0]{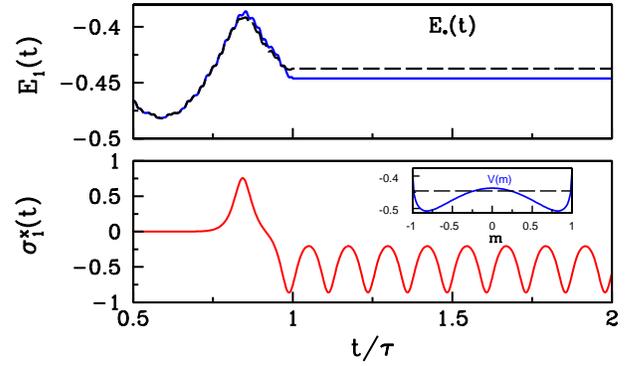}}
\vspace{-1.4cm}
\caption{Top panel: time evolution of $E_1(t)$, blue solid curve, and $E_*(t)=-N_1(t)\,h_1$, black dashed one, for $h_1=0.5$, $h_2=2$, $\omega=\Delta$, $E_0=0.1$ and the initial $T=1.1\,T_c$.  Bottom panel: corresponding evolution of the order parameter $\sigma^x_1(t)$. In the inset we show the classical potential after $t=\tau$ as well as the energy $E_1(\tau)$, dashed horizontal line.} 
\label{order}
\end{figure}

\begin{figure}[t]
\vspace{-0.8cm}
\centerline{\includegraphics[width=0.4\textwidth,angle=-90]{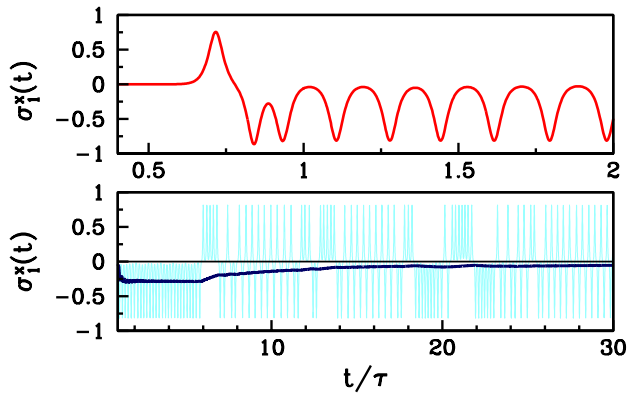}}
\vspace{-1cm}
\caption{Top panel: same as in Fig.~\ref{order} but in presence of a very small coupling $\lambda=10^{-2}$ between the two subsystems. Bottom panel: evolution of the order parameter for longer times. The solid dark blue line is the time-average value of $\sigma^x_1(t)$. } 
\label{orderbis}
\end{figure}

In view of the above results, we can conceive the possibility to start with subsystem 1 in the disordered phase above $T_c$ and end up, after the pulse, in its symmetry broken phase. When the perturbation is turned off, the subsystem 1 evolves unitarily with the Hamiltonian $H_1$ in \eqn{Hn}, which is equivalent to the classical motion in a potential\cite{Sengupta-PRB2006,Bapst&Semerjian-2012,Giacomo-PRB2012,Sciolla&Biroli-PRB2013} 
\be
V(m) = - \fract{N_1(\tau)^2}{2}\, m^2 - h_1\,N_1(\tau)\,\sqrt{1-m^2\;}\;,\label{V(m)}
\ee 
where $m = \langle \sigma^x_{1\bR} \rangle \in [-1,1]$ is the order parameter. The necessary condition for symmetry breaking to occur is that the potential \eqn{V(m)} has a double well, which requires $N_1(\tau)>h_1$. This is indeed well possible, see Fig.~\ref{Nt}. However, this is not sufficient; one also needs the conserved energy $E_1(\tau)$ to be lower than the top of the barrier separating the two minima 
$E_*(\tau) = V(0) = - N_1(\tau)\,h_1$, the broken-symmetry edge of Ref.~\cite{Giacomo-PRB2012}. If  this indeed happens, then the system will end up after the pulse into one of the two equivalent wells and keep oscillating around the minimum, which would imply a finite time-average value of the order parameter.  
In Fig.~\ref{order} we show an explicit case where that occurs, despite the initial temperature being greater than $T_c$, $T=1.1 \,T_c$. 
In the top panel we plot 
$E_1(t)$, blue solid curve, and $E_*(t)$, dashed black one. Indeed, at the end of the pulse the system has 
$E_1(\tau)<E_*(\tau)$. In the bottom panel we show the time evolution of the order parameter $\sigma^x_1(t)$, which, after 
the pulse, is found to oscillate in the left well with negative average value. \\
In real materials low- and high-energy degrees of freedom, here represented by subsystems 1 and 
2, respectively, are never decoupled from each other before and after the pulse, as we assume with model \eqn{Ham}. Therefore, some time after the pulse, the longer the weaker the coupling is, the excess energy acquired by subsystem 2 must flow back to 1 till the two equilibrate to a thermal stationary state. Such thermalisation never occurs with mean-field 
Hamiltonians like \eqn{Ham}. Nonetheless, we may wonder how the previous results change if already in the unperturbed Hamiltonian the two subsystems are coupled to each other. To that end, we 
changed\cite{Supplemental}  
$H$ of \eqn{Ham} into $H -\lambda\sum_\bR\,\sigma^x_{1\bR}\,\sigma^x_{2\bR}$, with a very small 
time-independent $\lambda=0.01$. The equilibrium phase diagram is practically unchanged by such a tiny $\lambda$ with respect to Fig.~\ref{phase-diagram}, with the only difference that now the finite 
$\langle \sigma^x_{1\bR}\rangle$ induces also a finite $\langle \sigma^x_{2\bR}\rangle \sim 
\lambda\,\langle \sigma^x_{1\bR}\rangle/h_2\ll 1$. In Fig.~\ref{orderbis} we show 
the time evolution of the order parameter $\sigma^x_1(t)$ with  the same Hamiltonian parameters of 
Fig.~\ref{order} but now with $\lambda\not= 0$. We note that shortly after the pulse end, top panel, the behaviour is  the same as for $\lambda=0$, bottom panel in Fig.~\ref{order}. However, for longer times, the small but finite $\lambda$ 
starts playing a role, energy flows back to subsystem 1 and eventually the order parameter vanishes on average, bottom panel in Fig.~\ref{orderbis}. In spite of that,  there does exist a sizeable time interval after the pulse where the system looks as it were ordered, albeit $T>T_c$.  \\
We finally remark that the circumstance that the system gets trapped after the pulse into one of the symmetry variant subspaces depends critically on $\tau$. This is evident from  
Fig.~\ref{order}, showing that a smaller $\tau$, below the peak in $E_1(t)$ at $t\sim 0.8\,\tau$, would not lead to the same result. We find that the trapping occurs most likely when the pulse stops just after the system, in its semiclassical motion, has jumped from one well into the other, which requires a sudden increase in energy to surpass the barrier and seems to be followed by $E_1(t)$ overshooting $E_*(t)$. We believe that in more realistic models endowed with dissipative channels that provide frictional forces to the semiclassical motion, such transient trapping into an ordered phase would still occur and not require too much fine-tuning of the pulse parameters.

In conclusion, we have shown in an exactly solvable toy model that a properly designed impulse perturbation can produce  an effective cooling of low-energy degrees of freedom at the expenses of high-energy ones that heat up. 
The model is extremely simple, but the mechanism is so generic that it may work even in more realistic situations, like in the case of photo-excited alkali-doped K$_3$C$_{60}$ fullerides as suggested in Ref.~\cite{mio}. 

We are grateful to Erio Tosatti and Alessandro Silva for useful comments and discussions. This work has been supported by 
the European Union under H2020 Framework Programs, ERC Advanced Grant No. 692670 ``FIRSTORM''.


    



\end{document}


\centerline{\Huge\textbf{Supplemental material}}
\bigskip
\bigskip
\bigskip
In this supplemental material we shall present in details the derivation of the variational density matrix quoted in the main text, the equations of motion that we integrated numerically to obtain the results shown in the figures, as well as some additional results in the case in which the two subsystems are coupled to each other even in the unperturbed Hamiltonian. 

\section{Variational approach to the infinitely-connected quantum Ising model}
We consider the Hamiltonian of an infinitely connected Ising model
\be
H = -\fract{1}{V}\,\sum_{\bR\bRp}\,\sigma^x_\bR\,\sigma^x_\bRp
-h\,\sum_\bR\,\sigma^z_\bR\,,\label{Ham}
\ee
where $\sigma^a_\bR$, $a=x,y,z$, are Ising operators defined at site $\bR$ of a lattice with $V=2M$ sites. We assume the following ansatz for the density matrix 
\be
\rho = \prod_\bR\,\rho_\bR\,,
\ee
with $\Tr\big(\rho_{n\bR}\big)=1$, together with the variational principle
\be
F \leq \min_{\{\rho_{\bR}\}}\bigg(\Tr\Big(\rho\,H\Big) 
-T\,\sum_\bR\,\Tr\Big(\rho_{\bR}\,\ln\rho_{\bR}\Big)\bigg)\,.\label{variational-principle}
\ee
Because of infinite connectivity, in the thermodynamic limit $M\to\infty$ the above inequality turns into a strict equality. 
Without loss of generality, we can take for $\rho_\bR$ the density matrix of a spin in a dimensionless magnetic field 
\be
\mathbf{B} = B\,\Big(\sin\theta\,\cos\phi,
\sin\theta\,\sin\phi,\cos\theta\Big)\,,
\ee
which we thus write as 
\be
\rho_\bR = \fract{1}{\,2\cosh B\, }\; U_\bR\,\exp\Big(B\,
\sigma^z_\bR\Big)\,U_\bR^\dagger\,,
\ee
where 
\be
U_\bR = \exp\bigg(-i\fract{\phi}{2}\,\sigma^z_{\bR}\bigg)
\exp\bigg(-i\fract{\theta}{2}\,\sigma^y_{\bR}\bigg)\,.
\ee
We note that, because of the factorised form of the density matrix,  
\beal
\Tr\Big(\rho\,\sigma^x_\bR\Big) &= 
\Tr\Big(\rho_\bR\,\sigma^x_\bR\Big) = 
\Tr\Bigg(
\fract{\text{e}^{\,B\,\sigma^z_\bR}}{2\cosh B}\,
\Big(\cos\theta\,\cos\phi\,\sigma^x_{\bR} - \sin\phi\,\sigma^y_{\bR} +\sin\theta\,\cos\phi\,
 \sigma^z_{\bR}\Big)\Bigg)\\
 &= \sin\theta\,\cos\phi\,
 \Tr\Bigg(
\fract{\text{e}^{\,B\sigma^z_\bR}}{2\cosh B}\;
 \sigma^z_{\bR}\Bigg) = \sin\theta\,\cos\phi\,\tanh B\,,
 \\
\Tr\Big(\rho\,\sigma^y_\bR\Big) &= 
\Tr\Big(\rho_\bR\,\sigma^y_\bR\Big) = 
\Tr\Bigg(
\fract{\text{e}^{\,B\sigma^z_\bR}}{2\cosh B}\,
\Big(\cos\theta\,\sin\phi\,\sigma^x_{\bR} + \cos\phi\,\sigma^y_{\bR} +\sin\theta\,\sin\phi\,
 \sigma^z_{\bR}\Big)\Bigg)\\
 &= \sin\theta\,\sin\phi\,
 \Tr\Bigg(
\fract{\text{e}^{\,B\,\sigma^z_\bR}}{2\cosh B}\;
 \sigma^z_{\bR}\Bigg) = \sin\theta\,\sin\phi\,\tanh B\,,
 \\ 
 \Tr\Big(\rho\,\sigma^z_\bR\Big) &= 
\Tr\Big(\rho_\bR\,\sigma^z_\bR\Big) = 
\Tr\Bigg(
\fract{\text{e}^{\,B\,\sigma^z_\bR}}{2\cosh B}\,
\Big(-\sin\theta \sigma^x_\bR +\cos\theta
 \sigma^z_{\bR}\Big)\Bigg)\\
 &= \cos\theta\,
 \Tr\Bigg(
\fract{\text{e}^{\,B\,\sigma^z_\bR}}{2\cosh B}\;
 \sigma^z_{\bR}\Bigg) = \cos\theta\,\tanh B\,.
\eal
We define 
\be
N \equiv \tanh B \;\Longrightarrow\; B = \fract{1}{2}\;
\ln\fract{1+N}{1-N}\;.
\ee
It follows that the internal energy per site is
\beal
E(N,\theta,\phi) &= \fract{1}{V}\,\Tr\Big(\rho\,H\Big) = 
- \fract{1}{2}\,N^2\;\sin^2\theta\,\cos^2\phi 
- h\,N\;\cos\theta\,,
\eal
while the entropy per site reads
\beal
S(N,\theta,\phi) &= -\fract{1}{V}\,\sum_\bR\,\Tr\Big(\rho_{\bR}\,\ln\rho_{\bR}\Big) = -\fract{1}{V}\,\sum_\bR\,\Tr\Bigg[
\fract{\text{e}^{\,B\,\sigma^z_\bR}}{2\cosh B}
\bigg(B\,\sigma^z_\bR - \ln 2\cosh B\bigg)\Bigg]\\
&= -B\,\tanh B + \ln 2\cosh B\\
&= -\fract{1+N}{2}\,\ln \fract{1+N}{2}
-\fract{1-N}{2}\,\ln \fract{1-N}{2}\;\equiv S(N).
\eal
The variational free energy per site is therefore 
\be
F(N,\theta,\phi) = E(N,\theta,\phi) - TS(N)\,,
\ee
whose minimisation leads to the results presented in the main text.
In particular we mention that the density matrix that minimises the free energy is such that 
\beal
U_\bR\,\exp\Big(B\,
\sigma^z_\bR\Big)\,U_\bR^\dagger &= \exp\bigg[
-\beta\,\Big(-\sin\theta(T)\,\sigma^x_\bR -
h\,\sigma^z_\bR\Big)\bigg] \equiv 
\exp\Big(-\beta\,H_\bR(T)\Big)\,,\label{H-eq}
\eal
where 
\be
\sin^2\theta(T) = 1 - \fract{h^2}{N(T)^2}\;,
\ee
with $N(T)$ defined through
\be
N(T) = \tanh\fract{N(T)}{T}\;,
\ee
for $h<1$ and 
\be
T \leq T_c = \fract{2h}{\ln\fract{1+h}{1-h}}\;,
\ee
otherwise  
\be
N(T)= \tanh\fract{h}{T}\;.
\ee
  \\

Finally we show that, despite the above density matrix does not commute with the total spin operator, in the thermodynamic limit its relative fluctuation does vanish. The total spin operator is defined through 
\beal
\mathbf{S}\cdot\mathbf{S} &=
\fract{1}{4}\,\sum_{\bR\bRp}\,
\boldsymbol{\sigma}_\bR\cdot\boldsymbol{\sigma}_\bRp
= \fract{3}{4}\,V + 
\fract{1}{4}\,\sum_{\bR\not =\bRp}\,
\boldsymbol{\sigma}_\bR\cdot\boldsymbol{\sigma}_\bRp\,.
\eal
Its expectation value is therefore
\beal
\Tr\Big(\rho\,\mathbf{S}\cdot\mathbf{S}\Big) &= 
\fract{3}{4}\,V + \fract{V(V-1)}{4}\,N^2
= M^2\,N^2 + \fract{M}{2}\,\Big(3-N^2\Big)\;
\underset{M\to\infty}{\longrightarrow} \; M^2\,N^2\,.
\eal
One readily finds that 
\beal
\Tr\bigg(\rho\,\Big(\mathbf{S}\cdot\mathbf{S}\Big)^2\bigg) &=
\fract{1}{16}\,\sum_{\bR_1 \bR_2 \bR_3 \bR_4}\,
\Tr\bigg(\rho\,
\Big(\boldsymbol{\sigma}_{\bR_1}\cdot\boldsymbol{\sigma}_{\bR_2}\Big)
\;\Big(\boldsymbol{\sigma}_{\bR_3}\cdot\boldsymbol{\sigma}_{\bR_4}
\Big)\bigg)\\
&= M^4\,N^4 + M^3\,N^2\,\Big(5-3N^2\Big) + O\big(M^2\big)\,,
\eal
so that 
\beal
\sqrt{\Tr\bigg(\rho\,\Big(\mathbf{S}\cdot\mathbf{S}\Big)^2\bigg)
- \Tr\Big(\rho\,\mathbf{S}\cdot\mathbf{S}\Big)^2\;} 
= M^{\frac{3}{2}}\,N\,\sqrt{2\,\Big(1-N^2\Big)\;}\,,
\eal
implying that the relative fluctuation vanishes as $1/\sqrt{M}$ in the thermodynamic limit.

\section{Equations of motion}
The time-dependent Hamiltonian that we consider in the main text is 
defined as 
\beal
H(t) &= -\sum_{n=1}^2 \Bigg(
\fract{1}{2}\sum_{\bR\bRp}\,\sigma^x_{n\bR}\,\sigma^x_{n\bRp}
+ h_n\,\sum_\bR\,\sigma^z_{n\bR}\Bigg)
-E(t)\,\sum_\bR\,\sigma^x_{1\bR}\,\sigma^x_{2\bR}\,,
\eal
where $t\geq 0$ and 
\be
E(t) = 
\begin{cases}
E_0\,\cos\omega t & t\leq \tau\,,\\
0 & t > \tau\,.
\end{cases}
\ee
We can write the density matrix still in a factorised form as 
\be
\rho(t) = \prod_\bR\,\rho_\bR(t)\,,
\ee
where $\rho_\bR(t)$ satisfies the equation of motion ($\hbar=1$)  
\be
i\,\dot{\rho}_\bR(t) = \Big[H_\bR(t)\,,\,\rho_\bR(t)\Big]\,,
\ee
with
\beal
H_\bR(t) &= -\sum_{n=1}^2\,\Big(m_n(t)\,\sigma^x_{n\bR}
+ h_n\,\sigma^z_{n\bR} \Big)
- E(t)\,\sigma^x_{1\bR}\,\sigma^x_{2\bR}\,.
\eal
The field $m_n(t)$ is determined self-consistently through 
\be
m_n(t) = \fract{1}{V}\,\sum_\bR\,\Tr\Big(\rho_\bR(t)\,\sigma^x_{n\bR}\Big) = \Tr\Big(\rho_\bR(t)\,\sigma^x_{n\bR}\Big)\;,\forall\,\bR\,.
\label{SC}
\ee
The boundary condition at $t=0$ is 
\be
\rho_\bR(t=0) = \rho_{1\bR}(T)\,\rho_{2\bR}(T)\,,
\ee 
where 
\be
\rho_{n\bR}(T) = \fract{\text{e}^{-\beta H_{n\bR}(T)}}
{\Tr\Big(\text{e}^{-\beta H_{n\bR}(T)}\Big)}\;,
\ee
with $H_{n\bR}(T)$ defined as in Eq.~\eqn{H-eq} for each Ising species, so that $H_\bR(t=0) = H_{1\bR}(T) + H_{2\bR}(T)$. \\
We define
\beal
\sigma^a_n(t) &= \fract{1}{V}\,\sum_\bR\,\Tr\Big(\rho_\bR(t)\,\sigma^a_{n\bR}\Big) = \Tr\Big(\rho_\bR(t)\,\sigma^a_{n\bR}\Big)\,,\\
\Sigma^{ab}(t) &= \fract{1}{V}\,\sum_\bR\,\Tr\Big(\rho_\bR(t)\,\sigma^a_{1\bR}\,\sigma^b_{2\bR}\Big) = \Tr\Big(\rho_\bR(t)\,\sigma^a_{1\bR}\,\sigma^b_{2\bR}\Big)\,,
\eal
which satisfy the equations of motion 
\beal
\dot{\sigma}^a_n(t) &= -i \,
\Tr\bigg(\rho_\bR(t)\,\Big[\sigma^a_{n\bR}\,,\,H_\bR(t)\Big]\Big)
\,,\\
\dot{\Sigma}^{ab}(t) &=  
-i\,\Tr\bigg(\rho_\bR(t)\,\Big[\sigma^a_{1\bR}\,\sigma^b_{2\bR}
\,,\,H_\bR(t)\Big]\bigg)\,.
\eal
By means of Eq.~\eqn{SC} we find 
\beal
\dot{\sigma}^x_1(t) &= 2h_1\,\sigma^y_1(t)\,,\\
\dot{\sigma}^y_1(t) &= -2h_1\,\sigma^x_1(t) + 
2\sigma^x_1(t)\,\sigma^z_1(t)
+ 2E(t)\,\Sigma^{zx}(t)\,,\\
\dot{\sigma}^z_1(t) &= -2\sigma^x_1(t)\,\sigma^y_1(t)
- 2E(t)\,\Sigma^{yx}(t)\,,\\
&\\
\dot{\sigma}^x_2(t) &= 2h_2\,\sigma^y_2(t)\,,\\
\dot{\sigma}^y_2(t) &= -2h_2\,\sigma^x_2(t) + 
2\sigma^x_2(t)\,\sigma^z_2(t)
+ 2E(t)\,\Sigma^{xz}(t)\,,\\
\dot{\sigma}^z_2(t) &= -2\sigma^x_2(t)\,\sigma^y_2(t)
- 2E(t)\,\Sigma^{xy}(t)\,,\label{equations-1}
\eal
as well as
 \beal
\dot{\Sigma}^{xx}(t) &= 2h_1\,\Sigma^{yx}(t) + 2h_2\,\Sigma^{xy}(t)
\,,\\ 
\dot{\Sigma}^{xy}(t) &= 2h_1\,\Sigma^{yy}(t) - 2h_2\,\Sigma^{xx}(t) +2\sigma^x_2(t)\,
\Sigma^{xz}(t) +2E(t)\,\sigma^z_2(t)
\,,\\
\dot{\Sigma}^{xz}(t) &= 2h_1\,\Sigma^{yz}(t) -2\sigma^x_2(t)\,\Sigma^{xy}(t) -2E(t)\,\sigma^y_2(t)
\,,\\
&\\
\dot{\Sigma}^{yx}(t) &=  -2h_1\,\Sigma^{xx}(t) + 2h_2\,\Sigma^{yy}(t) 
+ 2\sigma^x_1(t)\,\Sigma^{zx}(t) + 2E(t)\,\sigma^z_1(t) 
\,,\\
\dot{\Sigma}^{yy}(t) &=  -2h_1\,\Sigma^{xy}(t) -2h_2\,\Sigma^{yx}(t) 
+2\sigma^x_1(t)\,\Sigma^{zy}(t) + 2\sigma^x_2(t)\,\Sigma^{yz}(t)
\,,\\
\dot{\Sigma}^{yz}(t) &= -2h_1\,\Sigma^{xz}(t) + 
2\sigma^x_1(t)\,\Sigma^{zz}(t)
-2\sigma^x_2(t)\,\Sigma^{yy}(t) 
\,,\\
&\\
\dot{\Sigma}^{zx}(t) &= 2h_2\,\Sigma^{zy}(t) -2\sigma^x_1(t)\,\Sigma^{yx}(t) -2E(t)\,\sigma^y_1
\,,\\
\dot{\Sigma}^{zy}(t) &= -2h_2\,\Sigma^{zx}(t)
 -2\sigma^x_1(t)\,\Sigma^{yy}(t) + 2\sigma^x_2(t)\,\Sigma^{zz}(t)
\,,\\
\dot{\Sigma}^{zz}(t) &= -2\sigma^x_1(t)\,\Sigma^{yz}(t)
 -2\sigma^x_2(t)\,\Sigma^{zy}(t) 
\,,\label{equations}
\eal
with boundary conditions
\beal
\sigma^x_n(0) &= N_n(t)\,\sin\theta_n(T)\,,\\
\sigma^y_n(0) & = 0\,,\\
\sigma^z_n(0) &= N_n(T)\,\cos\theta_n(T))\,,\\
\Sigma^{ab}(0) &= \sigma^a_1(0)\,\sigma^b_2(0)\,.\label{BC}
\eal
We integrated numerically the set of equations \eqn{equations} with the initial conditions \eqn{BC} using the RKSUITE software package.~\cite{Rksuite} 

\section{Modified unperturbed Hamiltonian}

In the main text we mostly considered the case in which the unperturbed Hamiltonian is that of two decoupled infinitely-connected quantum Ising models, which are temporarily coupled to each other 
as long as the perturbation is on. This is a simpler situation easier to interpret. At the end, however, we showed few results obtained in the case in which the unperturbed Hamiltonian includes a coupling, specifically 
\beal
H \to H = -\sum_{n=1}^2\,\bigg(\fract{1}{2}\sum_{\bR\bRp}\,\sigma^x_{n\bR}\,\sigma^x_{n\bRp}
+ h_n\,\sum_\bR\,\sigma^z_{n\bR} \bigg)
- \lambda\,\sum_\bR\,\sigma^x_{1\bR}\,\sigma^x_{2\bR}\,.
\eal
The time evolution in the presence of the perturbation and with $\lambda\not=0$ are the same as those in equations  \eqn{equations-1} and \eqn{equations} with the only difference that  
\be
E(t) = 
\begin{cases}
\lambda + E_0\cos\omega t & 0<t\leq \tau\,,\\
\lambda & t>\tau\,.
\end{cases}
\ee
What changes are the boundary conditions at $t=0$. We need therefore to study the equilibrium state at $E_0=0$. The density matrix can be written as 
\be
\rho = \prod_\bR\,\rho_\bR\,,
\ee
where 
\be
\rho_\bR = \fract{\text{e}^{-\beta H_\bR}}{\Tr\Big(\text{e}^{-\beta H_\bR}\Big)}\;.
\ee 
We can choose 
\beal
H_\bR = -\sum_{n=1}^2\bigg(m_n(T)\,\sigma^x_{n\bR} + 
h_n\,\sigma^z_{n\bR}\bigg) - \lambda\,\sigma^x_{1\bR}\,\sigma^x_{2\bR}\,,\label{HR-2}
\eal
with the self-consistency condition 
\be
m_n(T) = \Tr\Big(\rho_\bR\,\sigma^x_{n\bR}\Big)\,,\quad \forall\,\bR\,.
\ee
In the main text we just focus on the state above $T_c$, hence we can assume that $m_n(T)=0$. The Hamiltonian \eqn{HR-2} can be easily brought to a diagonal form
\be
H_\bR = -\sum_{n=1}^2\,\bar h_n\,\bar\sigma^z_{n\bR}\,,
\label{HR-2-dia}
\ee
where, upon defining 
\beal
h &= h_1+h_2\,,\qquad \ep = h_2 - h_1\,,
\eal
and 
\be
\bar h = \sqrt{h^2+\lambda^2\;}\;,\qquad
\bar \ep = \sqrt{\ep^2+\lambda^2\;}\;,
\ee
then 
\be
\bar h_1 = \fract{\bar h -\bar\ep}{2}\;,\qquad 
\bar h_2 = \fract{\bar h +\bar\ep}{2}\;.
\ee 
The equilibrium state is therefore characterised by 
\beal
\langle\, \bar\sigma^z_{1\bR}\,\rangle &= \tanh\fract{\bar h_1}{T}
\equiv \tau_1\;,
\\
\langle\, \bar\sigma^z_{2\bR}\,\rangle &= \tanh\fract{\bar h_2}{T}
\equiv \tau_2\;,
\\
\langle\, \bar\sigma^{x(y)}_{n\bR}\,\rangle &= 0\,.
\label{BC-2-dia}
\eal

The original operators are related to those in Eq.~\eqn{HR-2-dia} 
through the following transformations:
\beal
\sigma^z_{1\bR} &= \bigg(\fract{h}{2\bar h} + 
\fract{\ep}{2\bar \ep}\bigg)\,\bar\sigma^z_{1\bR}
+ \bigg(\fract{h}{2\bar h} - 
\fract{\ep}{2\bar \ep}\bigg)\,\bar\sigma^z_{2\bR}\,,\\
\sigma^z_{2\bR} &= \bigg(\fract{h}{2\bar h} - 
\fract{\ep}{2\bar \ep}\bigg)\,\bar\sigma^z_{1\bR}
+ \bigg(\fract{h}{2\bar h} + 
\fract{\ep}{2\bar \ep}\bigg)\,\bar\sigma^z_{2\bR}\,,\\
\sigma^x_{1\bR} &= \cos\theta_-\;\bar\sigma^x_{1\bR} 
+\sin\theta_-\;\bar\sigma^z_{1\bR}\,\bar\sigma^x_{2\bR}\,,
\phantom{\bigg(\fract{h}{h}}\\
\sigma^x_{2\bR} &= \sin\theta_+\;\bar\sigma^x_{1\bR}\,
\bar\sigma^z_{2\bR} 
+\cos\theta_+\;\bar\sigma^x_{2\bR}\,,\label{transfo-2}
\eal
where 
\be
\cos 2\theta_\pm = \fract{h\ep \mp\lambda^2}{\bar h\bar\ep} \,.
\ee 
Eqs.~\eqn{transfo-2} and \eqn{BC-2-dia} allow to readily 
determine the initial conditions in the equations of motion
\beal
\sigma^z_1(0) &= \fract{h}{2\bar h}\;\big(\tau_1+\tau_2\big)
- \fract{\ep}{2\bar \ep}\;\big(\tau_2-\tau_1\big)\,,\\
\sigma^z_2(0) &= \fract{h}{2\bar h}\;\big(\tau_1+\tau_2\big)
+ \fract{\ep}{2\bar \ep}\;\big(\tau_2-\tau_1\big)\,,\\
\Sigma^{xx}(0) &= \lambda\,\Bigg[
\fract{1}{2\bar h}\;\big(\tau_1+\tau_2\big) 
+\fract{1}{2\bar\ep}\;\big(\tau_2-\tau_1\big)\Bigg]\,,\\
\Sigma^{yy}(0) &= \lambda\,\Bigg[
-\fract{1}{2\bar h}\;\big(\tau_1+\tau_2\big) 
+\fract{1}{2\bar\ep}\;\big(\tau_2-\tau_1\big)\Bigg]\,, \\
\Sigma^{zz}(0) &= \tau_1\,\tau_2\,,
\eal
all others being zero. \\

Starting from the disordered phase, one can also calculate the critical temperature for the appearance of spontaneous symmetry breaking. Suppose we perturb the Hamiltonian with 
\be
\delta H = -\sum_{n=1}^2\,\sum_\bR\, m_n\,\sigma^x_{n\bR}\,.
\ee
In linear response, and upon defining the thermodynamic susceptibilities 
\be
\chi_{nm}(T) = -i\int_0^\infty dt\, \langle\,
\Big[\sigma^x_{n\bR}(t)\,,\,\sigma^x_{m\bR}\Big]\,\rangle\,,
\ee
which can be readily calculated 
\beal
\chi_{11}(T) &= \fract{1}{2\bar h_1\,\bar h_2}\,\bigg[
\Big(\tau_1\,\bar h_2 
+ \tau_2\,\bar h_1\Big) + \fract{h\ep + \lambda^2}{\bar h\,\bar\ep}\,
\Big(\tau_1\,\bar h_2 - \tau_2\,\bar h_1\Big)\bigg]\,,\\
\chi_{22}(T) &= \fract{1}{2\bar h_1\,\bar h_2}\,\bigg[
\Big(\tau_1\,\bar h_2 +\tau_2\,\bar h_1\Big)
+ \fract{h\ep - \lambda^2}{\bar h\,\bar\ep}\,
\Big(\tau_2\,\bar h_1 -\tau_1\,\bar h_2\Big)\bigg]
\;,\\
\chi_{12}(T) &=\chi_{21}(T) = \tau_1\,\tau_2\,\fract{\lambda}{\bar h_1\,\bar h_2}\;,
\eal
we would find that 
\be
\langle \sigma^x_{n\bR}\rangle = 
\sum_{p=1}^2\,\chi_{np}(T)\;m_p\,.
\ee
Symmetry breaking occurs at a temperature $T=T_c$ at which we can solve the above equations with 
$m_n = \langle \sigma^x_{n\bR}\rangle$, which corresponds to
\be
\begin{pmatrix}
1 - \chi_{11}(T_c) & -\chi_{12}(T_c)\\
-\chi_{21}(T_c) & 1-\chi_{22}(T_c)
\end{pmatrix}\,
\begin{pmatrix}
m_1\\
m_2
\end{pmatrix} =0\,,
\ee 
namely to the condition 
\be
\Big(1 - \chi_{11}(T_c)\Big)\,
\Big(1 - \chi_{22}(T_c)\Big) = \chi_{12}(T_c)^2\,.\label{Tc-2}
\ee

